\documentclass[aps,prd,amsmath,amssymb,showpacs,showkeys,nofootinbib]{revtex4}
\usepackage{graphicx}
\usepackage{dcolumn}
\usepackage{bm}
\usepackage{epsfig}

\begin{document}
\title{Hadron form factors from sum rules for vacuum-to-hadron correlators}
\author{Dmitri Melikhov}
\affiliation{
SINP, Moscow State University, 119991, Moscow, Russia \\ 
HEPHY, Austria Academy of Sciences, Nikolsdorfergasse 18, A-1050, Vienna, Austria}
\date{\today}
\begin{abstract}
We analyse the extraction of the bound-state form factor from vacuum-to-hadron correlator, 
which is the basic object for the calculation of hadron form factors in the method of 
light-cone sum rules in QCD. 
We study this correlator in quantum mechanics, calculate it exactly, 
and derive the corresponding OPE. We then apply the standard procedures of 
QCD sum rules to isolate the ground-state form factor from this correlator. 
We demonstrate that fixing the effective continuum threshold, one of the key
ingredients of the sum-rule calculation of bound-state parameters, 
poses a serious problem for sum rules based on vacuum-to-hadron correlators. 
\end{abstract}
\pacs{11.55.Hx, 12.38.Lg, 03.65.Ge}
\keywords{Nonperturbative QCD, hadron properties, QCD sum rules}
\maketitle
\section{Introduction}
A calculation of hadron parameters from sum rules in QCD \cite{svz} involves two steps: 
(i) one calculates the operator product expansion (OPE) series for a relevant correlator and 
obtains the sum rule which relates this OPE to the sum over hadronic states, and (ii) 
one attempts to get rid of the contribution of the excited states to the sum 
and isolate the ground-state contribution, thus relating parameters 
of the ground state to parameters of QCD. 

Constructing OPE in QCD is a well-defined procedure, whereas the extraction of the ground-state 
contribution requires several assumptions which do not come from the underlying field theory;  
it is hard to control the reliability of these assumptions and, as the consequence, 
the accuracy of the ground-state parameters extracted from sum rules. 

In order to probe the reliability and the accuracy of standard procedures adopted in the method 
of sum rules in QCD, quantum-mechanical model is an ideal tool:\footnote{For a discussion of many 
aspects of sum rules in quantum mechanics we refer to \cite{nsvz,nsvz1,qmsr,orsay}.} 
in this model one can exactly calculate direct analogues of the field-theory correlators, 
one can generate analogues of the OPE for these correlators to any order, 
and, finally, one knows the exact bound-state parameters, such as masses, wave functions, and form factors. 
Therefore, one can compare the outcome of the standard sum-rule calculation with the exact values and in 
this way obtain an ``unbiased'' probe of the reliability of the sum-rule calculation. 

In the recent publications \cite{lms_sr,lms_gamma}, the procedures of extracting ground-state parameters 
from two-point \cite{svz} and three-point \cite{ioffe} vacuum correlators were studied. 
In the present paper, a similar strategy is applied to vacuum-to-hadron correlator \cite{stern} 
which is the basic object for the extraction of hadron form factors in the method of 
light-cone sum rules in QCD. 

We are going to demonstrate that fixing the effective continuum threshold for the case of vacuum-to-hadron 
correlator poses a challenging problem (in fact more challenging problem 
than that for the case of the vacuum-to-vacuum correlators). In particular, the effective continuum threshold 
is found to differ sizeably from nearly constant effective continuum threshold for polarization operator 
and to depend strongly both on the momentum transfer and on the Borel parameter.
This leads to difficulties in finding a reasonable approximation to 
this quantity and, as the consequence, to large systematic uncertainties in sum-rule results for hadron 
form factors. 


\section{The model} 
We consider a non-relativistic model with HO potential
\begin{eqnarray}
H=H_0+V(r), \quad H_0=\vec p^2/2m, \quad V(r)={m\omega^2 r^2}/{2}, \qquad r=|\vec r|.
\end{eqnarray}
The full Green function $G(E)=(H-E)^{-1}$ and the free Green function $G_0(E)=(H_0-E)^{-1}$ 
satisfy the equation 
\begin{eqnarray}
G^{-1}(E)-G_0^{-1}(E)=V, 
\end{eqnarray}
which may be solved by constructing the expansion in powers of the interaction $V$:
\begin{eqnarray}
\label{ls} G(E)=G_0(E)-G_0(E)VG_0(E)+\cdots.
\end{eqnarray}
In HO model all characteristics of the bound states are easily calculable, 
e.g. for the ground state one has 
\begin{eqnarray}
\label{E0} 
E_{\rm ground}=\frac{3}{2}\omega,\qquad 
\Psi_{\rm ground}(\vec r)=\left(\frac{m\omega}{\pi}\right)^{3/4}\exp\left({-\frac12 m\omega r^2}\right),
\qquad
F_{\rm ground}(q)=\exp(-q^2/4m\omega). 
\end{eqnarray}
The elastic form factor of the ground state $F_{\rm ground}(q)$ is defined according to 
\begin{eqnarray}
F_{\rm ground}(q)=\langle \Psi_{\rm ground}|J(\vec q)|\Psi_{\rm ground}\rangle=
\int d\vec k \psi_{\rm ground}^\dagger(\vec k)\psi_{\rm ground}(\vec k-\vec q)=
\int d\vec r |\psi_{\rm ground}(\vec r)|^2e^{i\vec q\vec r},\qquad q\equiv |\vec q|
\end{eqnarray}
with the current operator given by the kernel 
\begin{eqnarray}
\label{J}
\langle \vec r'|J(\vec q)|\vec r\rangle=\exp(i\vec q\vec r)\delta(\vec r-\vec r').
\end{eqnarray}


\section{The polarization operator}
The polarization operator  
\begin{eqnarray}
\label{pi}
\Pi(T)=\langle \vec r_f=0|\exp(- H T)|\vec r_i=0\rangle 
\end{eqnarray}
is used in the method of sum rules for the extraction of the wave function at the
origin (the decay constant) of the ground state \cite{svz}.  
A detailed discussion of the corresponding procedure for HO model was presented in \cite{lms_sr}. 
For HO potential, the analytic expression for $\Pi(T)$ is known \cite{nsvz}:
\begin{eqnarray}
\label{piexact}
\Pi(T)=\left(\frac{\omega m}{\pi}\right)^{3/2}\frac1{\left[2\sinh(\omega T)\right]^{3/2}}
\end{eqnarray}
The corresponding average energy is defined as follows
\begin{eqnarray}
\label{energypi}
E_\Pi(T)\equiv -\partial_T\log \Pi(T)=\frac32 \omega \coth(\omega T).
\end{eqnarray}
The OPE is the expansion of the polarization operator at small Euclidean time $T$ 
(or, equivalently, in powers of $\omega$):  
\begin{eqnarray}
\label{piope}
\Pi_{\rm OPE}(T)=\left(\frac{m}{2\pi T}\right)^{3/2}
\left[1-\frac{1}{4}\omega^2T^2+\frac{19}{480}{\omega^4 T^4}
+\cdots \right].  
\end{eqnarray}


\section{The vertex function}

The basic quantity for the extraction of the form factor 
in the method of dispersive sum rules is the correlator of three currents \cite{ioffe}. 
The analogue of this quantity in quantum mechanics has the form 
\begin{eqnarray}
\Gamma(\tau_2,\tau_1,q)=
\langle \vec r_f=0|\exp(-H \tau_2)J(\vec q)\exp(-H \tau_1)|\vec r_i=0\rangle,  
\end{eqnarray}
with $J(\vec q)$ defined in (\ref{J}). For equal times $\tau_1=\tau_2=T/2$, 
we obtained an explicit expression for HO model \cite{lms_gamma}
\begin{eqnarray}
\label{gamma}
\Gamma(T,q)=\left(\frac{m\omega}{\pi}\right)^{3/2}
\frac{1}{\left[2 \sinh(\omega T)\right]^{3/2}}\exp\left(-\frac{q^2}{4m\omega}
\tanh\left(\omega T/2\right)\right). 
\end{eqnarray}
The corresponding average energy has a simple form 
\begin{eqnarray}
\label{energygamma}
E_\Gamma(T,q)\equiv -\partial_T\log \Gamma(T,q)=\frac32 \omega \coth(\omega T)
+\frac{q^2}{4m}\frac{1}{\left(1+\cosh(\omega T)\right)}. 
\end{eqnarray}
The OPE is the expansion of $\Gamma(T,q)$ in powers of $\omega$;  
for its explicit form in HO model we refer to \cite{lms_gamma}. An unpleasant feature of $\Gamma_{\rm OPE}$ 
is the polynomial growth of power corrections with $q^2$ which limits the convergence of the truncated 
$\Gamma_{\rm OPE}$ to not very large $q^2$. Respectively, this correlator may be used for extracting form 
factors at not very large $q^2$. 

\section{The vacuum-to-hadron correlator}
To apply the method of sum rules to hadron form factors in a wider range of $q^2$, 
a different correlator is used -- the vacuum-to-hadron amplitude of the T-product of two currents. 
The analogue of this quantity in quantum mechanics has the form 
\begin{eqnarray}
A(E,q)=\langle \vec r=0|G(E)J(\vec q)|\Psi_{\rm ground}\rangle, 
\end{eqnarray}
or, after Borelization, 
\begin{eqnarray}
\label{A}
A(T,q)=\langle \vec r=0|G(T)J(\vec q)|\Psi_{\rm ground}\rangle. 
\end{eqnarray}
An obvious disadvantage of this correlator compared to the three-point function 
$\Gamma$ is the necessity to know the ground-state wave function for its calculation. 
As we shall see, as a bonus, this correlator has an enhanced contribution of the ground 
state which makes it potentially more attractive for the extraction of the ground-state form factor. 
   
In HO model, the ground state wave function is known explicitly. Making use of this 
wave function (\ref{E0}) and the results for the Green function in 
configuration space $G(r,T)=\langle \vec r'=0|G(T)|\vec r\rangle$ from \cite{nsvz}, 
we obtain the exact analytic expression for $A$ in HO model: 
\begin{eqnarray}
\label{Aexact}
A(T,q)=\left(\frac{m\omega}{\pi}\right)^{3/4}\exp\left(-\frac{3}{2}\omega T\right)
\exp\left(-\frac{q^2}{4m\omega}(1-e^{-2\omega T})\right). 
\end{eqnarray}
The function $A(T,q)$ depends on two dimensionless variables ${q^2}/{4m\omega}$ and $\omega T$. 
The average energy for this correlator has a simple form 
\begin{eqnarray}
\label{Eexact}
E(T,q)\equiv -\partial_T \log A(T,q)=\frac32\omega+\frac{q^2}{2m}\exp(-2\omega T). 
\end{eqnarray}
It should be noticed that the properties of the correlator $A(T,q)$ differ considerably from the properties 
of $\Pi(T)$ and $\Gamma(T,q)$: for instance, $A(T,q)$ and the corresponding $E(T,q)$ are finite at $T=0$, 
whereas the correlators $\Pi$ and $\Gamma$ diverge at small $T$ as $T^{-3/2}$.  
The energies, related to $\Pi$ and $\Gamma$, $E_{\Pi}$ and $E_{\Gamma}$, diverge as $1/T$. 

\subsection{Ground-state contribution to $A(T,q)$}
The ground-state contribution to the correlator reads 
\begin{eqnarray}
A_{\rm ground}(T,q)=\psi_{\rm ground}(r=0)\exp\left({-E_{\rm ground} T}\right)F_{\rm ground}(q) 
\end{eqnarray}
leading to  
\begin{eqnarray}
\label{Rground}
\frac{A_{\rm ground}(T,q)}{A(T,q)}=\exp\left(-\frac{q^2}{4m\omega}e^{-2\omega T}\right). 
\end{eqnarray}
Notice the following features: 

\noindent
(i) At large $T$, the ground state provides the dominant contribution to $A$, similar to any other correlator:  
\begin{eqnarray}
A(T,q)\to \Psi_{\rm ground}(r=0)e^{-E_{\rm ground}T}F_{\rm ground}(q)+\ldots
\end{eqnarray}
(ii)
At small $q$, the ground state dominates the correlator for all $T$. 
This is a specific feature of $A$ which arises due to the choice of the initial state. 
Thus, compared with $\Pi(T)$ and $\Gamma(T,q)$, the correlator $A$ ``maximizes'' the ground state contribution.  

\subsection{OPE for $A(T,q)$}
\begin{figure}
\includegraphics[width=10cm]{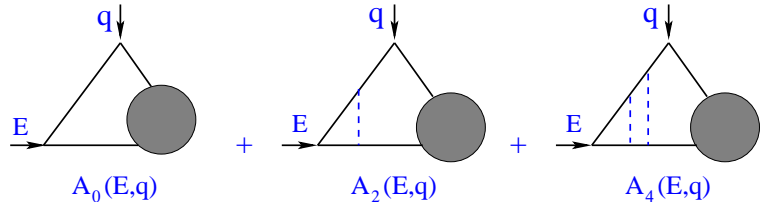}
\caption{\label{Fig:1} Expansion of the Green function $G(E)$ in the correlator $A(E,q)$ in 
powers of the interaction. Full blobs denote the wave function of the ground state, 
dashed lines correspond to the potential $V$.}
\end{figure}
In order to obtain the OPE for $A$ along the lines of QCD, one should expand the Green 
function $G$ in Eq.~(\ref{A}) in powers of the interaction (Fig.~\ref{Fig:1}). 
This generates the expansion of $A(T,q)$ in powers of $(\omega T)^2$.
To obtain this series from the exact expression requires however some care: one should take into 
account that the parameter $\omega$ enters both the wave function and the Green function $G$. 
Therefore, in practice it is convenient to proceed as follows: treat the parameter 
in $G$ as $\omega$, but the same parameter in the wave function as $\omega_0$; 
calculate the exact correlator $A(T,q|\omega,\omega_0)$ from Eq.~(\ref{A}); 
expand this expression in powers of $\omega$ and obtain 
$A(T,q|\omega,\omega_0)=\sum_{n=0}^\infty(\omega T)^{2n}A_{2n}(T,q|\omega_0)$; 
finally, set $\omega_0\to\omega$ in the functions $A_{2n}(T,q|\omega_0)$. 
This procedure yields a rather complicated OPE series: 
\begin{eqnarray}
A(T,q)=\sum\limits_{n=0}^\infty(\omega T)^{2n} A_{2n}(T,q|\omega), 
\end{eqnarray} 
where the functions $A_n(T,q|\omega)$ have a nontrivial $T$, $q$, and $\omega$-dependence:   
\begin{eqnarray}
A_0&=&\left(\frac{m\omega}{\pi}\right)^{3/4}\frac{1}{(1+\omega T)^{3/2}}
\exp\left(-\frac{q^2}{2m\omega}\frac{\omega T}{1+\omega T}\right), \nonumber\\
A_2&=&-\frac{1}{12}\left(\frac{m\omega}{\pi}\right)^{3/4}
\frac{1}{(1+\omega T)^{7/2}}
\exp\left(-\frac{q^2}{2m\omega}\frac{\omega T}{1+\omega T}\right)
\left[3(1+\omega T)(3+\omega T)-{2 q^2 T}/{ m} \right]. 
\end{eqnarray} 
For large $q^2$, one obtains 
\begin{eqnarray}
A_n\sim (q^2)^{n-1}\exp\left(-\frac{q^2}{2m\omega}\frac{\omega T}{1+\omega T}\right).
\end{eqnarray} 
Obviously, for large $q^2$ all functions $A_n$ remain finite and tend to zero, 
so that the truncated OPE may be calculated for any value of $q^2$.
However, $A_{n+2}/A_n\sim q^2$ and thus, for a purely confined potential, 
the truncated OPE {\it does not} reproduce the exact $A(T,q)$ in the 
limit $q^2\to\infty$.\footnote{Compare with the following example: 
imagine we expand $f=e^{-2x}=e^{-x}e^{-x}=(1-x+x^2/2+\ldots)e^{-x}\equiv \sum_{n=0}^\infty f_n(x)$.
For $x\to \infty$, $f_n(x)\to 0$, so any truncated series takes a finite value for $x\to \infty$.  
However, $f_{n+1}(x)/f_n(x)\to \infty$, and thus a truncated series does not reproduce 
the exact $f(x)$ for large $x$.}
\begin{figure}[b]
\begin{tabular}{cc}
\includegraphics[width=7cm]{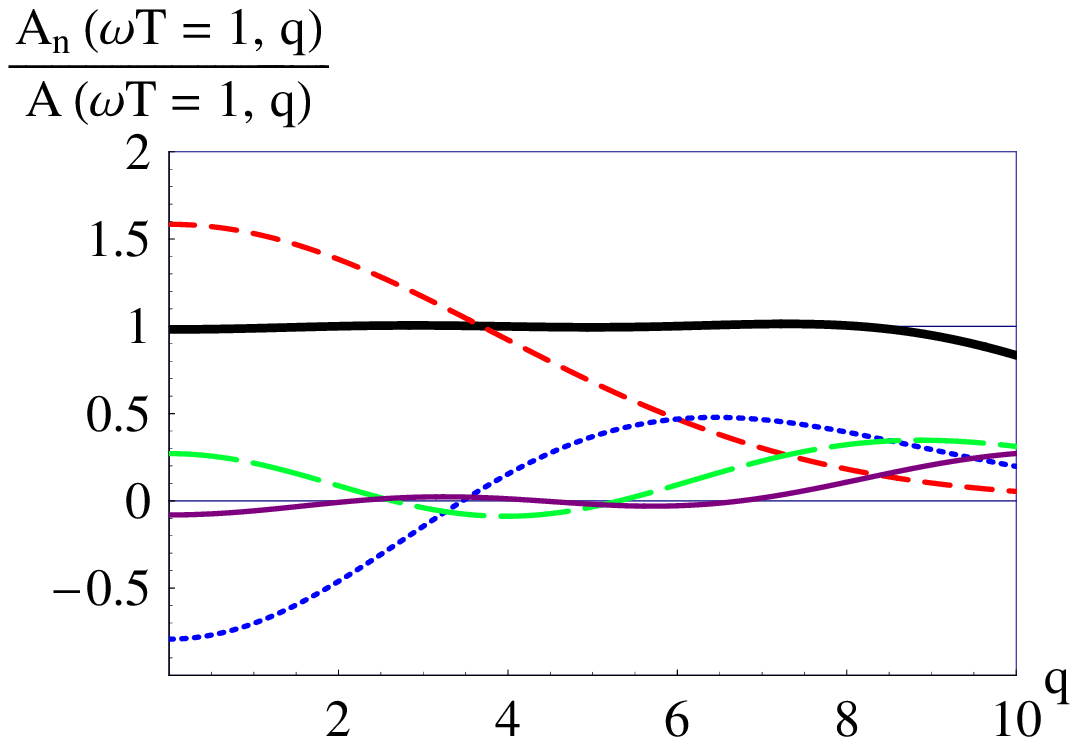}&
\includegraphics[width=7cm]{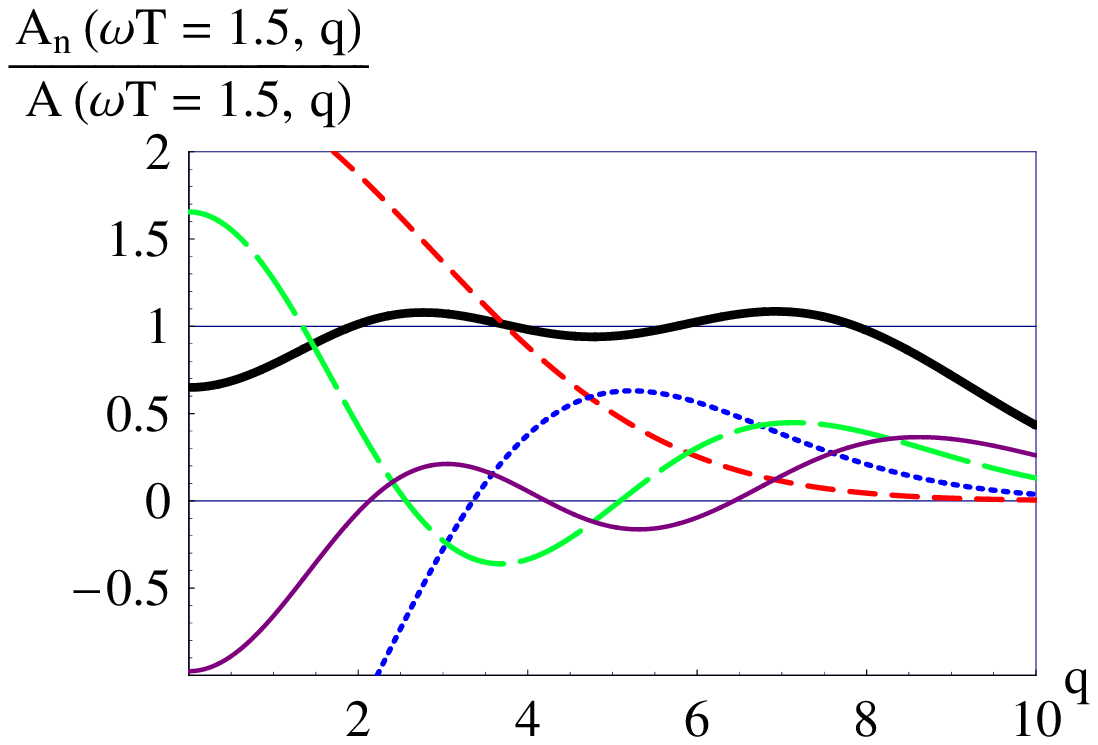}
\end{tabular}
\caption{\label{Plot:fixed_T}
The accuracy of the OPE series truncated to 4 terms 
$A_{\rm OPE, 4}\equiv \sum_{n=0}^3(\omega T)^{2n} A_{2n}$ 
for $A(T,q)$ vs $q$ for two different values of $\omega T$: 
(a) $\omega T=1.0$, (b) $\omega T=1.5$. 
Dashed (red) line - $A_0$, dotted (blue) - $A_2(\omega T)^2$, long-dashed (green) - $A_4(\omega T)^4$, 
solid (violet) - $A_6(\omega T)^6$, Thick solid (black) line - $A_{\rm OPE, 4}$.} 
\end{figure}
Fig.~\ref{Plot:fixed_T} illustrates the accuracy of the OPE series, truncated to 4 terms, 
$A_{\rm OPE, 4}\equiv \sum_{n=0}^3 (\omega T)^{2n}A_{2n}$. In order to have a 
good description of the exact function $A(T,q)$ for $q\le 8\omega$ with $A_{\rm OPE, 4}$ one 
should stay in the region $\omega T\le 1\div 1.2$. 


\subsection{Sum rule and exact effective continuum threshold}
The sum rule represents the equality of the correlator calculated in quark and in hadron basis:
\begin{eqnarray}
\label{sr}
A_0(T,q) +A_{\rm power} (T,q)=A_{\rm ground}(q)+A_{\rm excited}(T,q). 
\end{eqnarray}
According to the standard procedures of the method of sum rules, the contribution of the ground state 
is assumed to be dual to the low-energy region, i.e. the region of small momentum transfers. 
To implement this idea technically, one should represent the correlator as the dispersion representation 
in $z=\vec k^2$, and then cut the integral at the effective continuum threshold $z_{\rm eff}$. The 
easiest way to derive the spectral representation for $A(T,q)$ is to use the full Green function in 
momentum space
\begin{eqnarray}
G(\vec k^2,T)\equiv 
\langle \vec r=0|G(T)|\vec k\rangle=
\frac{1}{(2\pi)^{3/2}}\frac{1}{[\cosh(\omega T)]^{3/2}}e^{-\frac{\vec k^2}{2 m \omega}\tanh(\omega T)}. 
\end{eqnarray}
The correlator $A(T,q)$ takes the form  
\begin{eqnarray}
A(T,q)=\int d\vec k \;G(\vec k^2,T)\Psi_{\rm ground}(\vec k-\vec q)=\int_0^\infty dz e^{-z T}a(z,q)
\end{eqnarray}
The explicit expression for $a(z,q)$ may be easily obtained using $\Psi_{\rm ground}$ given in (\ref{E0}). 

The cut correlator is defined according to 
\begin{eqnarray}
\label{Acut}
A^{\rm cut}(T,q,z_{\rm eff})\equiv\int_{0}^{z_{\rm eff}} e^{-z T}a(z,q)dz,   
\end{eqnarray}
and it should reproduce the ground-state contribution: 
\begin{eqnarray}
\label{sr1}
\psi_{\rm ground}(r=0)e^{-{E_{\rm ground}}T}F_{\rm ground}(q)=A^{\rm cut}(T,q,z_{\rm eff}(T,q)).
\end{eqnarray}
This expression is precise if we use an exact $T$- and $q$-dependent effective continuum threshold 
(which cannot be calculated knowing only OPE, but can of course be reconstructed in HO model 
since we know the exact form factor, ground-state energy, and ground-state mass). 
Moreover, Eq.~(\ref{sr1}) 
may be understood as the definition of the exact effective continuum threshold, 
if one makes use of the exact hadron parameters on the l.h.s. 
Obviously, Eq.~(\ref{sr1}) alone is not predictive and the form factor (as well as any other parameter) 
of the ground state may be obtained in the method of sum rules only if one imposes an independent 
criterion to fix the effective continuum threshold. Recall that this criterion does not come directly from 
the underlying theory. 
\begin{figure}
\begin{tabular}{cc}
\includegraphics[width=6cm]{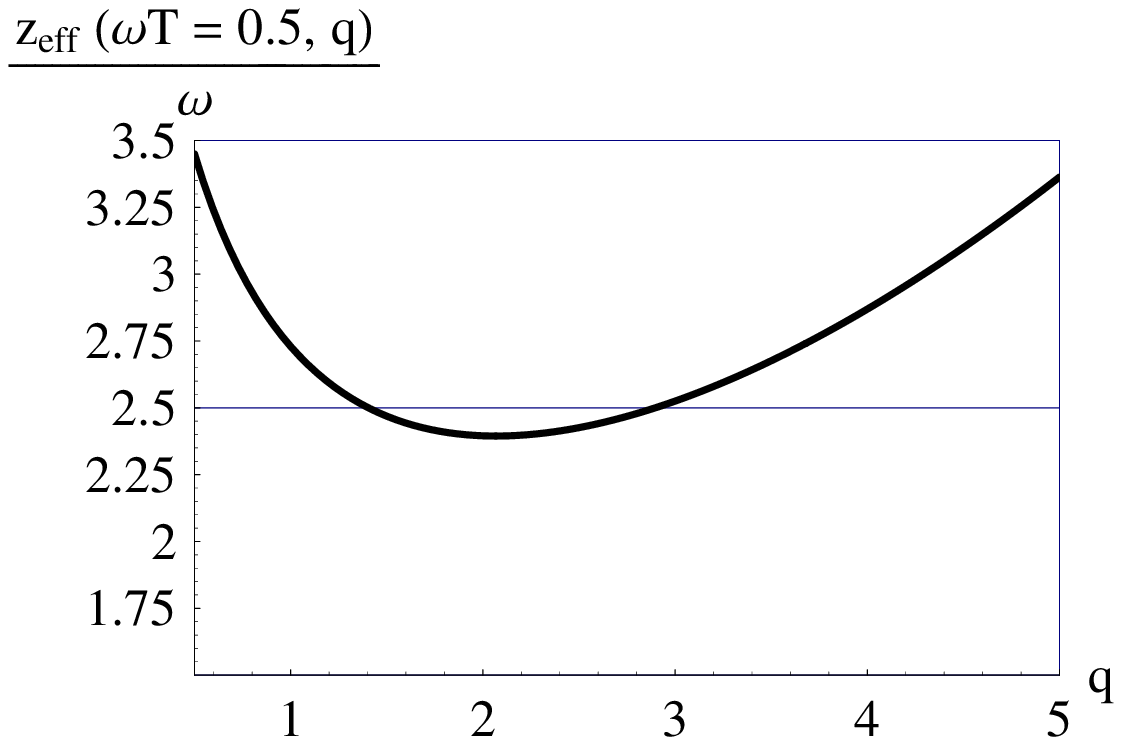}&
\includegraphics[width=6cm]{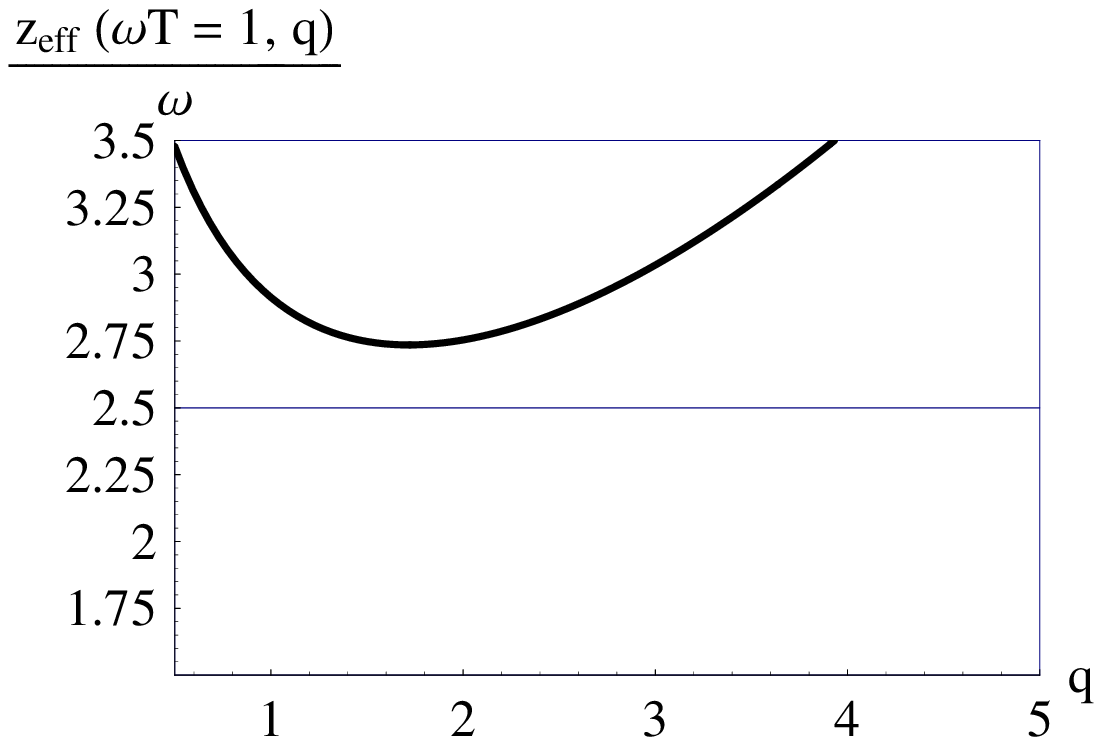}
\end{tabular}
\caption{\label{Plot:zeff}
The exact effective continuum threshold $z_{\rm eff}(T,q)$ [as obtained by solving (\ref{sr1}) for the 
exact bound-state parameters in the l.h.s.] vs $q$ for 
$\omega T=0.5$ (a) and $\omega T=1$ (b). 
A horizontal line at $2.5$ is given as a benchmark: this constant was shown to provide a good 
approximation to the exact continuum threshold in sum rule for $\Pi(T)$ \cite{lms_sr}.} 
\end{figure}
One may expect that in the case of the correlator $A(T,q)$, the exact $z_{\rm eff}$ strongly 
depends on the momentum transfer: e.g., as noticed above, the ground state dominates the correlator 
at small $q$, therefore $z_{\rm eff}(T,q)\to\infty$ as $q\to 0$. Fig.~(\ref{Plot:zeff}) shows 
$z_{\rm eff}(T,q)$ as obtained by solving (\ref{sr1}) for the exact 
parameters of the ground state from (\ref{E0}). Indeed, the exact effective continuum threshold 
strongly 
depends both on $T$ and $q$. Moreover, it is rather far from the value $z_c=2.5\,\omega$ 
which was found to provide a good approximation for the exact continuum threshold for 
two-point sum rule \cite{lms_sr}.

Recall that the effective continuum threshold is one of the two basic ingredients 
(along with the OPE for the correlator) which determine the value of the ground state parameter. 
Therefore, the observed strong dependence of the exact effective continuum threshold on 
$q$ and $T$ is 
very unpleasant: it implies difficulties in finding a reasonable approximation to this quantity and, 
as the consequence, large uncertainties in the sum-rule results for the hadron form factor.

\subsection{Extraction of the ground-state form factor}

Let us now apply the standard procedures and extract the form factor 
at two values of $q$: at the intermediate $q=2\omega$ ($q^2/4m\omega=1$) 
and at relatively large $q=4\omega$ ($q^2/4m\omega=4$). 

In realistic cases, the exact correlator is not known and one may calculate only several terms of 
its OPE series. Assume we have 4 terms at our disposal: 
\begin{eqnarray}
A_{\rm OPE, 4}=\sum_{n=0}^3(\omega T)^{2n} A_{2n}(T,q).
\end{eqnarray}
First, let us determine the ``window'' -- the region of the Borel parameter 
where, according to standard criteria, we may work to extract the ground-state 
form factor: 
To be sure that our truncate OPE provides a good description of the exact expression -- say, with 
1\% accuracy -- we must stay at not very large values of $T$. This requirement gives the upper 
boundary of the Borel ``window'', $\omega T\le 1.2$
The lower boundary of the ``window'' is determined by the requirement that 
the gound state gives a sizeable contribution -- say, more than 50\% -- 
to the correlator. This yields $0.25\le \omega T$ for $q=2\omega$ and 
$0.8\le \omega T$ for $q=4\omega$.

So the ``window'' is $0.25\le \omega T \le 1.2$ for $q=2\omega$ and 
$0.8\le \omega T \le 1.2$ for $q=4\omega$.\footnote{Notice 
that the ``window'' disappears for large values of $q$: with a fixed
number of terms in the truncated OPE, the upper boundary of the ``window'' decreases with $q$, 
whereas the lower boundary increases, as can be seen from (\ref{Rground}).}  

Second, we need to apply the cut to the correlator. Let us define the cut correlator in a 
bit different way than in the previous Section, to make it closer to the standard 
sum-rule calculation in QCD: namely, we apply the cut to $A_0$ and leave higher terms intact. 
Then the cut correlator has the form 
\begin{eqnarray}
\label{cutA0}
A^{\rm cut}(T,q,z_c)&=&A_0^{\rm cut}(T,q,z_c)+A_2(T,q)+A_4(T,q)+A_6(T,q),\quad
A^{\rm cut}_0(T,q,z_c)=\int_0^{z_c} dz \exp(-z T)a_0(z,q).\nonumber\\
\end{eqnarray}
Here $a_0$ is the spectral density of $A_0$: 
\begin{eqnarray}
A_0(T,q)=\int d\vec k G_0(\vec k^2,T)\Psi_{\rm ground}(\vec k-\vec q)=\int_0^\infty dz e^{-z T} a_0(z,q),
\qquad G_0(\vec k^2,T)=\frac{1}{(2\pi)^{3/2}}e^{-T\vec k^2/2m}.
\end{eqnarray}
The form factor is related to the cut correlator according to relation (\ref{sr1}). 

Next, the key step -- to impose a criterion for fixing $z_c$. 
In the literature, two prescriptions for fixing $z_c$ have been used:

\noindent
(A) One assumes $z_c$ to be a $T$- and $q$-independent constant and its value is chosen to 
be the same as in the two-point sum rule \cite{braun}. In our case this implies setting $z_c=2.5\,\omega$. 

 \noindent
(B) One assumes $z_c$ to be $T$-independent, but tunes its value for each value of $q$, separately. 
To this end one calculates the average energy of the cut correlator 
\begin{eqnarray}
E(T,q,z_c)\equiv -\partial_T \log A(T,q,z_c),
\end{eqnarray}
which depends on $T$ and $q$ because of the approximation $z_{\rm eff}(T,q)\to z_c$ (see discussion in \cite{lms_sr}). 
Then, one determines $z_c$ for each $q$ such that the function $E(T,q,z_c)$ 
has a horizontal tangent $E=E_{\rm ground}$ \cite{ball}.  

We also present the exact effective continuum threshold, which is obtained by solving Eq.~(\ref{sr1}) 
for the known ground-state parameters on its l.h.s., and the correlator with the cut applied according 
to (\ref{cutA0}) on its r.h.s.

\begin{figure}
\begin{tabular}{ccc}
\includegraphics[width=5.8cm]{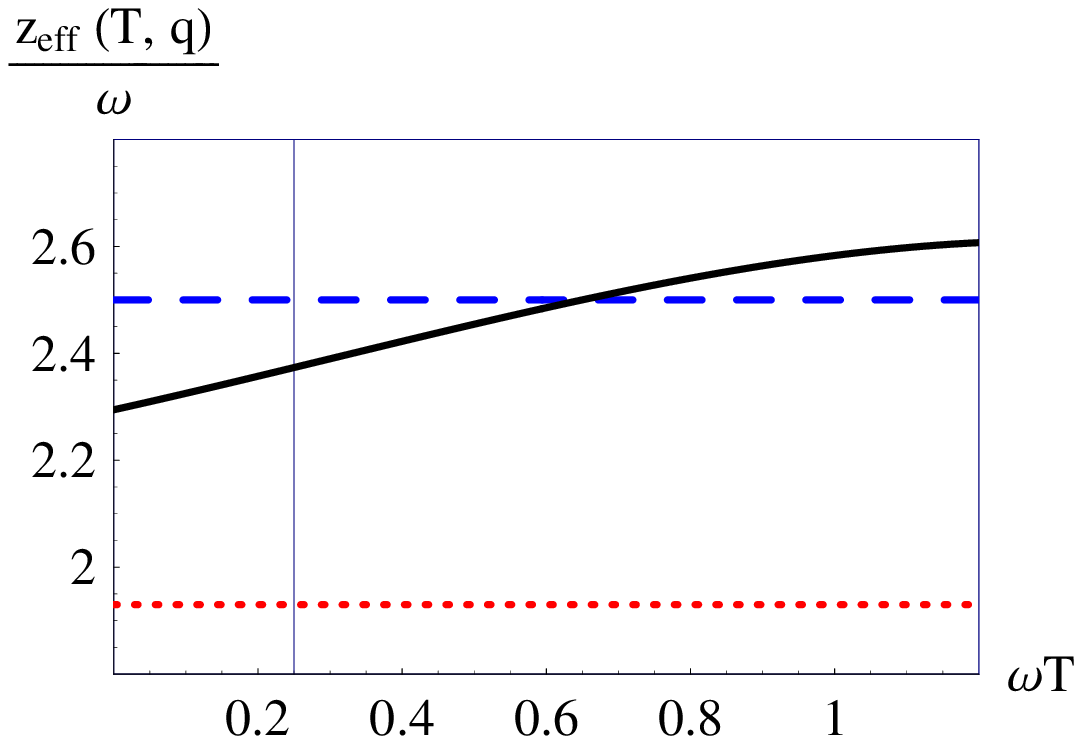}
&
\includegraphics[width=5.8cm]{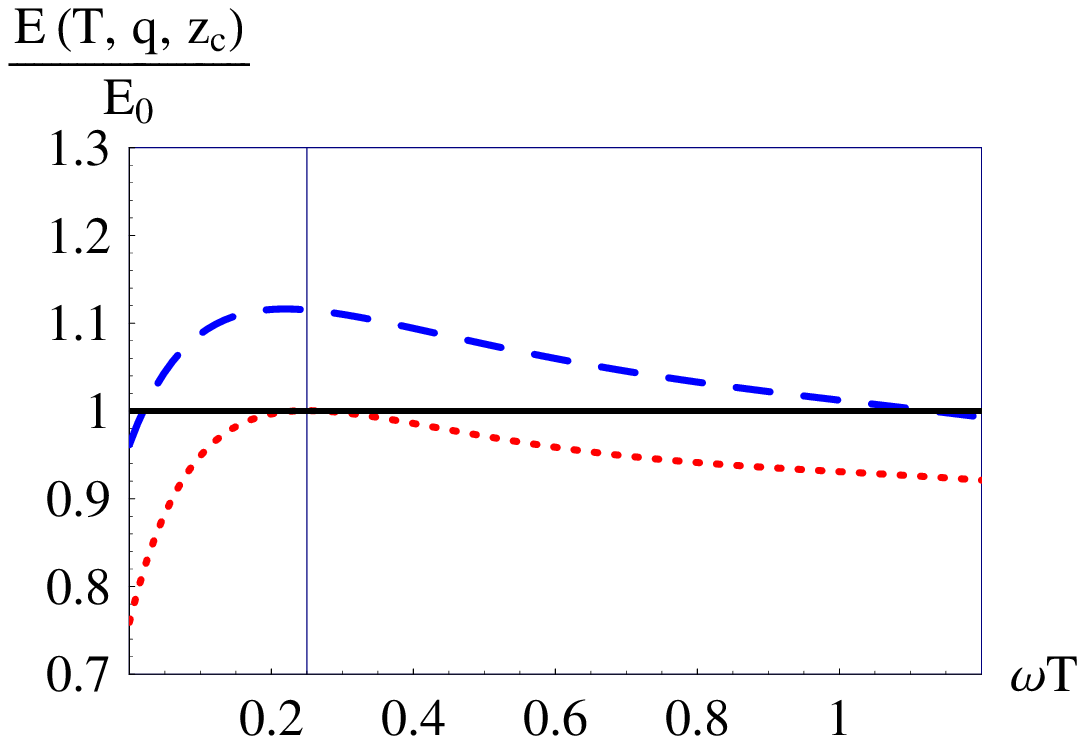}
&
\includegraphics[width=5.8cm]{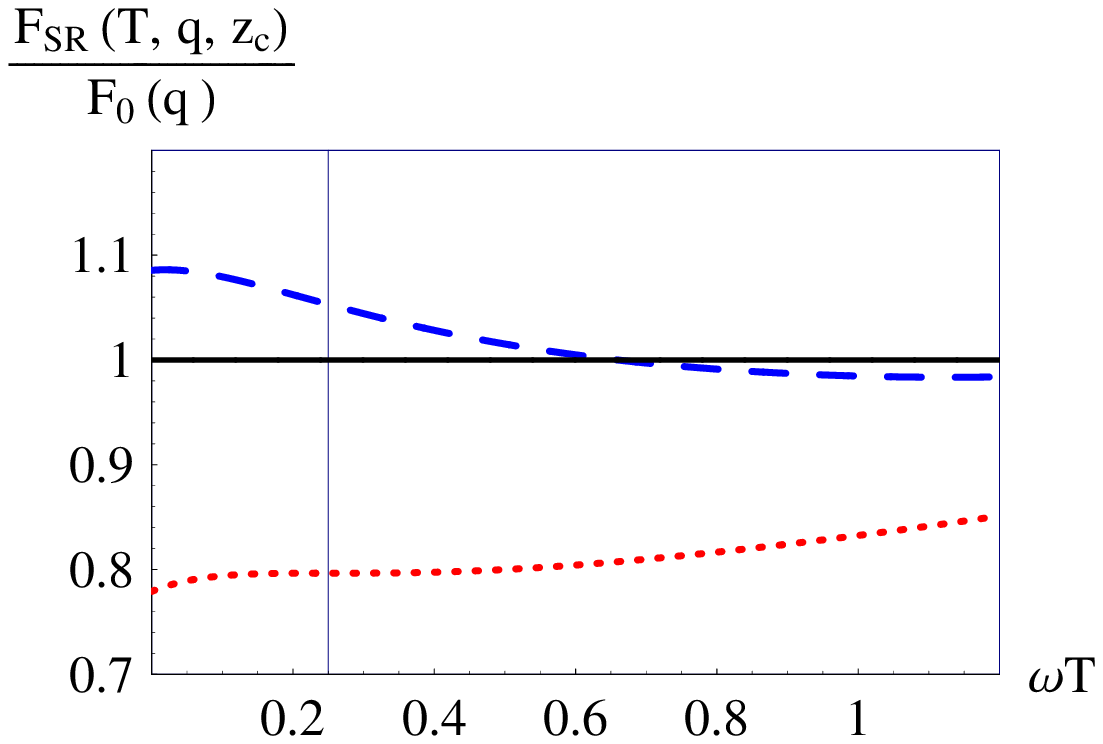}
\end{tabular}
\caption{\label{Plot:q=2}
Intermediate momentum transfer $q=2\omega$ ($q^2/4m\omega=1$). 
Dashed (blue) line - constant effective continuum threshold $z_c=2.5\,\omega$ (prescription A);
dotted (red) - $T$-independent effective continuum threshold tuned according to prescription B; 
solid (black)  - $T$-dependent exact effective continuum threshold $z_{\rm eff}(T,q)$.}
\end{figure}

\begin{figure}
\begin{tabular}{ccc}
\includegraphics[width=5.8cm]{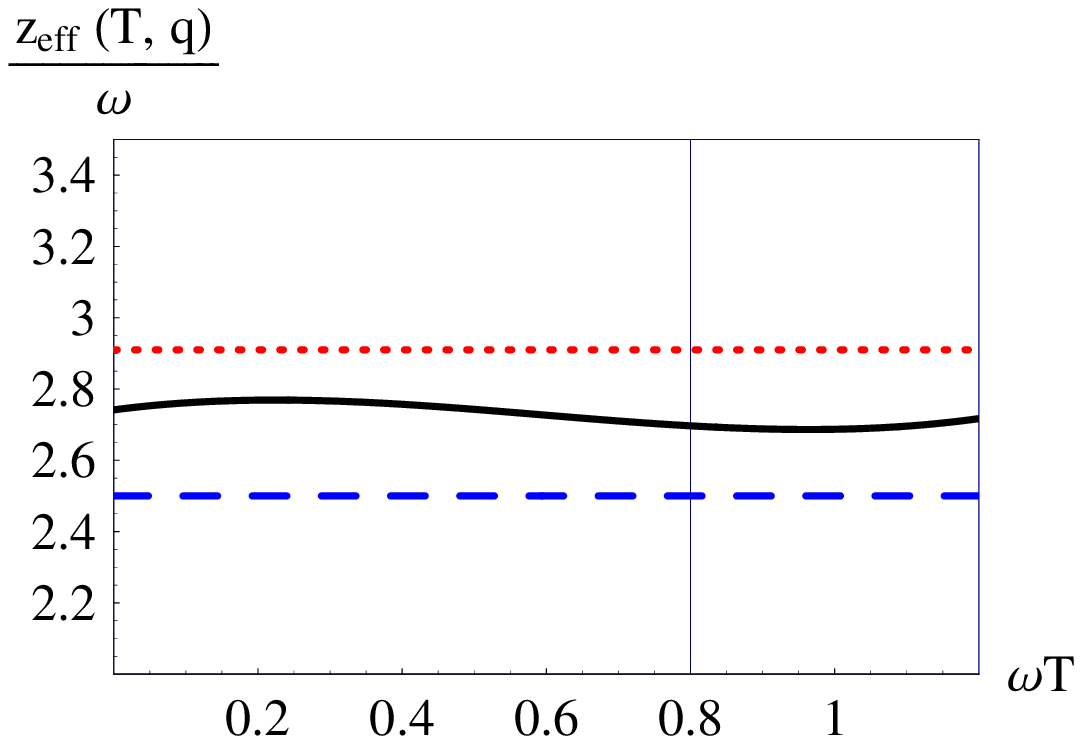}
&
\includegraphics[width=5.8cm]{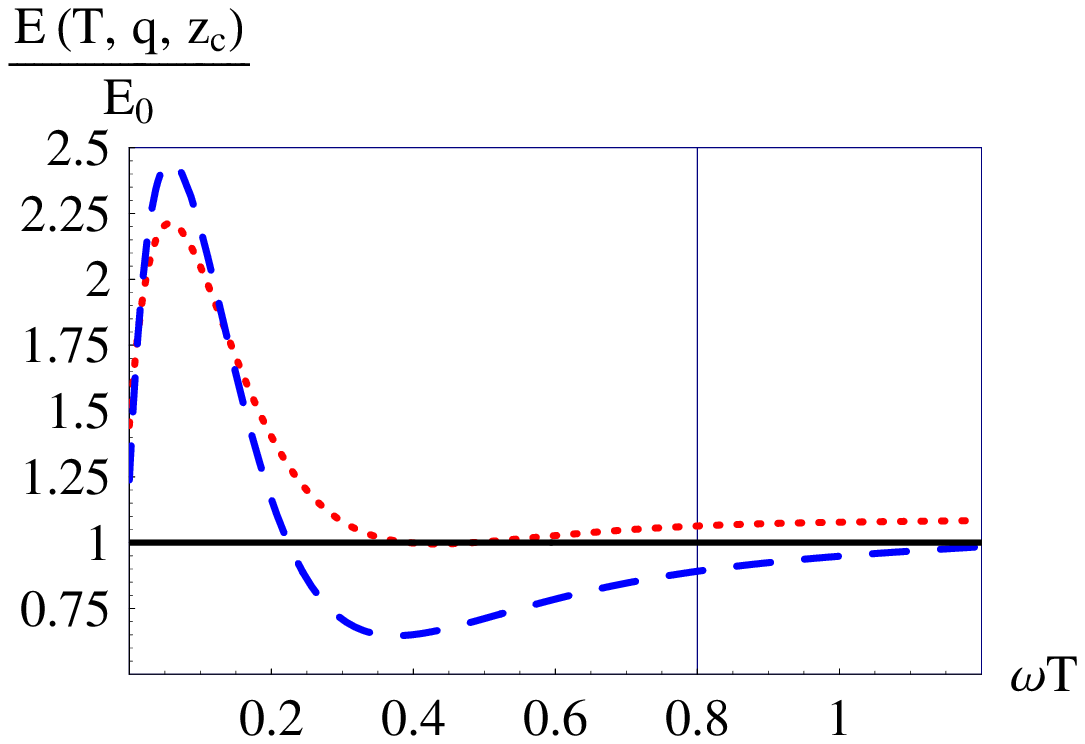}
&
\includegraphics[width=5.8cm]{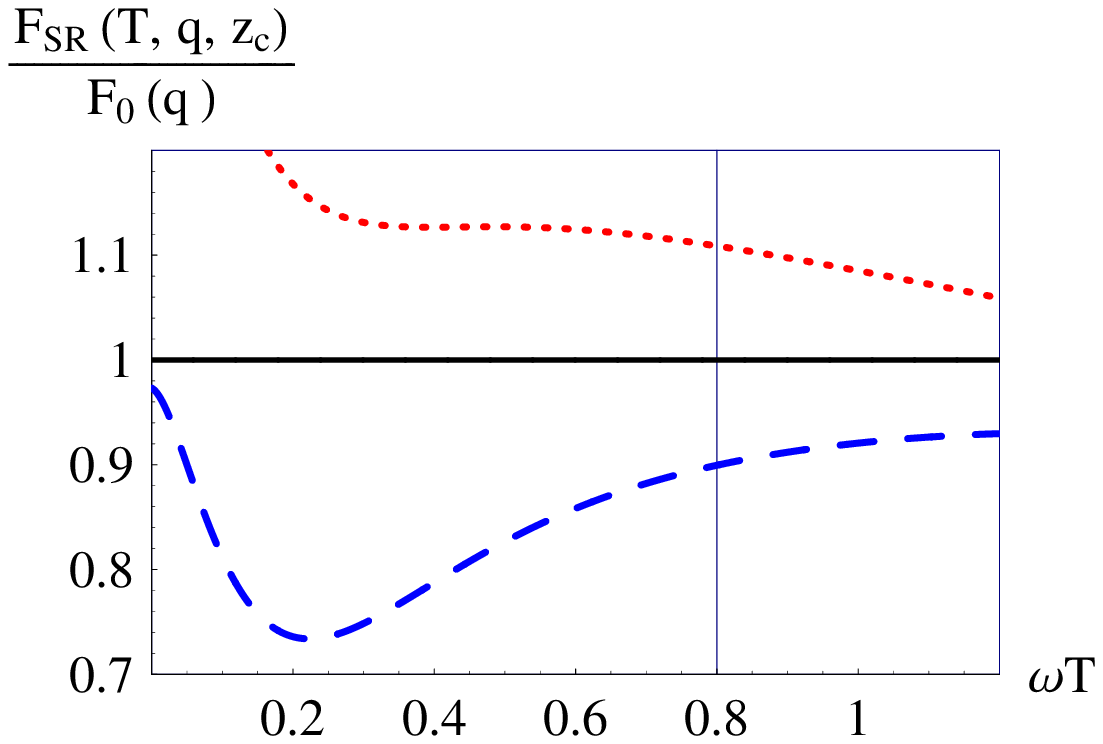}
\end{tabular}
\caption{\label{Plot:q=4}
Relatively large momentum transfer $q=4\omega$ ($q^2/4m\omega=4$).
Dashed (blue) line - constant effective continuum threshold $z_c=2.5\,\omega$ (prescription A);
dotted (red) - $T$-independent effective continuum threshold tuned according to prescription B; 
solid (black)  - $T$-dependent exact effective continuum threshold $z_{\rm eff}(T,q)$.}
\end{figure}

The corresponding plots are given in Figs.~\ref{Plot:q=2} and \ref{Plot:q=4}. 
According to the usual procedures, the theoretical uncertainty of the sum-rule prediction is obtained 
from the range covered by the form factor when varying the Borel parameter within the ``window''. 
 
For $q=2\omega$ (Fig.~\ref{Plot:q=2}) the exact effective threshold turns out to depend sizeably on $T$ 
in the ``window''. Prescription B leads to a stable form factor 
$F_{\rm SR}/F_{\rm ground}=0.82\pm 0.02$ which however underestimates the actual value by
almost 20\%; prescription A leads in this case to a very good estimate  
$F_{\rm SR}/F_{\rm ground}=1.02\pm 0.03$. 

For $q=4\omega$ (Fig.~\ref{Plot:q=4}) the situation is different:    
The exact continuum threshold is almost flat in the ``window'', however its value is different from the values
determined by both prescriptions A and B. As the result, the prescription A strongly underestimates the form
factor, leading to $F_{\rm SR}/F_{\rm ground}=0.91\pm 0.01$, whereas the prescription B overestimates it, yielding 
$F_{\rm SR}/F_{\rm ground}=1.08\pm 0.02$. 

The lesson to be learnt from these two examples is clear --  
neither the recepy A or B can guarantee the extraction of the true 
form factor with a controlled accuracy: in ``good''
cases one of these prescriptions may work, in ``bad'' cases they fail. 
Unfortunately, so far we have not found any criterion 
which could discriminate between the ``good'' and the ``bad'' cases. 


\section{Conclusions}
We studied the vacuum-to-hadron correlator $A(T,q)$ in harmonic-oscillator model and analysed 
the procedure of extracting the ground-state form factor from 
this correlator using the procedures of the method of sum rules. 
The great advantage of this model is the possibility to obtain {\it exact} expression for the correlator 
and the knowledge of {\it exact} parameters of the ground state. Therefore this model provides a 
brilliant opportunity to study the reliability of the procedures used in the method of sum rules.  

Our main conclusions are summarized below: 

\noindent 1. 
Compared to vacuum-to-vacuum correlator, the correlator $A(T,q)$ requires the knowledge of the 
ground-state wave function for its calculation. As a bonus, it contains an enhanced 
contribution of the ground state, which makes it potentially a 
favourable candidate for the calculation of the ground state form factor with sum rules. 

\noindent 2.
However, the extraction of the ground-state form factor from $A(T,q)$ faces a challenging problem of 
fixing the effective continuum threshold: as demonstrated in our exactly solvable model, the exact 
effective continuum threshold -- 
a solution of the sum rule for the known exact ground-state parameters -- strongly depends 
both on the Borel parameter $T$ and the momentum transfer $q$. 
This implies serious difficulties in finding a reasonable approximation to this quantity and, as the 
consequence, large systematic uncertainties in sum-rule results for hadron form factors. 

\noindent 3. 
We tested prescriptions used in the literature for fixing the effective continuum threshold and 
found that none of these prescriptions guarantees the reliable extraction of the form factor: for some 
values of the momentum transfer they may provide a good result, for other values of the momentum transfer 
they may fail. Unfortunately, so far we have not found any criterion 
which could discriminate between the ``good'' and the ``bad'' situations.  

\noindent 4. 
We have seen that the usual criteria adopted in the method of sum rules 
for obtaining error estimates of the ground-state form factors 
(Borel stability, variation of the form factor in the ``window'') 
do not guarantee a good extraction of the form factor and do not provide realistic error estimates --  
the actual errors turn out to be considerably larger. 
In this respect hadron form factors obtained from the vacuum-to-hadron correlators 
share the same difficulties as hadron parameters obtained from other versions of sum rules 
(see conclusions in \cite{lms_sr,lms_gamma}). 

A systematic study of the actual accuracy of hadron form factors obtained 
from vacuum-to-hadron correlators (in particular, with light-cone sum rules) should be undertaken 
before these form factors may be used in problems where rigorous error estimates are required. 

\vspace{.5cm}

\noindent
{\it Acknowledgments:}
I would like to thank Wolfgang Lucha, Silvano Simula, and Hagop Sazdjian for interesting discussion,  
and the Theory Group of the Institute of Nuclear Physics, University Paris-Sud, 
for hospitality during my stay in Orsay. 
Financial support from the Austrian Science Fund (FWF) under project P17692, 
RFBR under project 07-02-00551, and CNRS is gratefully acknowledged.  


\end{document}